\documentstyle[11pt,aaspp4]{article} 
\begin{document}


\noindent
NATURE 422, 871-873 (issue 24 April 2003)

\bigskip \bigskip 

\centerline{{\large\bf First-generation black-hole-forming supernovae and 
}}
\centerline{{\large\bf 
the metal abundance pattern of a very iron-poor star
}} 

\bigskip 

\bigskip \centerline{\large Hideyuki Umeda \& Ken'ichi Nomoto} 

\bigskip 

\centerline{Department of Astronomy, School of Science, 
University of Tokyo, Tokyo 113-0033, Japan} 
\centerline{Received 18 December 2002; Accepted 4 March 2003}



\bigskip 

\noindent{\bf
 It has been proposed$^1$ theoretically that the
first generation of stars in the Universe (population III) would be
as massive as 100 solar masses (100$M_\odot$),
because of inefficient cooling$^{2-4}$ of the precursor gas clouds.
Recently, the most iron-deficient (but still carbon-rich) low-mass 
star -- HE0107-5240 -- was discovered$^5$. If this is a population III 
that gained its metals (elements heavier than helium) after its formation, 
it would challenge 
the theoretical picture of the formation of the first stars. 
Here we report that the patterns of elemental abundance in
HE0107-5240 (and other extremely metal-poor stars) are in good accord with 
the nucleosynthesis that occurs in stars with
masses of $20-130M_\odot$ when they become supernovae if, during the explosions,
the ejecta undergo substantial mixing and fall-back to form massive black holes. 
Such supernovae have been observed$^7$. 
The abundance patterns are not, however,  
consistent with enrichment by supernovae from stars in the range $130-300 M_\odot$. 
We accordingly infer that the first-generation supernovae came 
mostly from explosions of $ \sim 20-130M_\odot$ stars; 
some of these produced iron-poor but carbon- and
oxygen-rich ejecta. Low-mass second-generation stars, like HE0107-5240, could
form because the carbon and oxygen provided pathways for gas to cool.}

\bigskip

 The abundance pattern of HE0107-5240 is characterized by very 
large C/Fe and N/Fe ratios,
while the abundances of elements heavier than Mg are 
as low as that of Fe (Fig. 1). How could this peculiar 
abundance pattern have be formed? We note that this is not 
the only extremely metal-poor (EMP) star to have large C/Fe and N/Fe ratios -- 
several other such stars 
have been discovered $^{8-10}$. Therefore a reasonable explanation of the abundance pattern 
should also apply to other EMP stars. 

 Before describing our model, we examine two alternative explanations of the 
abundance pattern of C-rich EMP stars$^5$. These explanations assume that small-mass stars 
formed from EMP gases, and that carbon was enhanced after the formation. The first is 
mass transfer from a companion star. This explanation suffers from 
two difficulties. (1) Not all EMP stars have
observational evidence for the existence of the existence of 
companion stars, though further observations to obtain radial velocity data
are necessary$^{11}$.  
(2) Some EMP stars (for example, CS 22949-037 and CS 29498-043) 
have a relatively large Si abundance, [Si/Fe] $\sim 1$ (refs 8-10). 
(Here [A/X] = 
log$_{10} (N_A / N_X) -$ log$_{10} (N_A / N_X ) _\odot$, where  
the subscript $\odot$ refers to the Sun and $N_A$ and $N_X$
are the number densities of elements A and X respectively.)
The overabundance of Si is very difficult to explain by 
mass transfer from companion AGB stars.

 The second explanation is matter accretion onto a population III star during repeated 
passages through the Galactic disk$^6$. This model alone cannot explain 
the large abundance of C and N -- for example, in HE0107-5240, with [C/H] = $-1.3$. 
Therefore, C and N 
need to be produced in the interior of the EMP stars. 
Recently, nucleosynthesis in the 
low-mass population III stars was studied extensively$^{12, 13}$. 
Such studies reveal that C and N 
abundances can be significantly enhanced if mixing is efficient 
during the He-flash in the core
at the tip of the red-giant branch. However, HE0107-5240 is 
identified as a low-mass $(\sim 0.8M_\odot$) red
giant$^5$, and thus this mechanism is not likely to operate.

 Here we consider a model in which C-rich EMP stars are produced in the ejecta of 
(almost) metal-free supernovae mixed with EMP interstellar matter. We 
use population III pre-supernova progenitor models$^{14}$, simulate the supernova explosion, and 
calculate detailed nucleosynthesis. The elemental abundances 
of one of our models are in good agreement with HE0107-5240 (Fig. 1). In this model, the 
progenitor mass is 25$M_\odot$ and the explosion energy E$_{\rm exp}$ = 
3 $\times 10^{50}$ erg (or E$_{51}$ = E$_{\rm exp}$ / 
10$^{51}$ = 0.3). The resultant abundance distribution is shown 
in Fig. 2. The processed material is assumed to mix 
uniformly in the region from $M_{\rm r}$ = 1.8 $M_\odot$ to 6.0 $M_\odot$ 
(where $M_{\rm r}$ denotes the lagrangian mass coordinate measured from
the centre of the pre-supernova model).
Such a large-scale mixing was 
found to take place in SN1987A and various explosion models$^{15, 16}$. 
Almost all materials 
below $M_{\rm r} = 6.0 M_\odot$ 
falls back to the central remnant and only a small fraction (factor $f$ = 
2 $\times 10^{-5}$) is ejected from this region. 

 The CNO elements in the ejecta were produced by pre-collapse He shell burning 
in the He layer, which contains $0.2M_\odot$ $^{12}$C. Mixing of H into the He shell-burning region 
produces $4 \times 10^{-4} M_\odot~ ^{14}$N. On the other hand, only a small amount of heavier elements 
(Mg, Ca, and Fe-peak elements) are ejected and their abundance ratios are the average 
over the region of $M_{\rm r} = 1.8 - 6.0 M_\odot$. 
The subsolar ratios of [Ti/Fe] = $-0.4$ and [Ni/Fe] =  
$-0.4$ are the results of the relatively small explosion energy (E$_{51}$ = 0.3). With this 
`mixing and fallback', the large C/Fe and C/Mg ratios observed in HE0107-5240 are 
well reproduced. In this model, N/Fe appears to be underproduced. However, N can be 
produced inside the EMP stars through the C-N cycle, and brought up to the surface 
during the first dredge up stage while becoming a red-giant star$^{17}$. 

The `mixing and fallback' is commonly required to reproduce the abundance 
pattern of typical EMP stars$^{14}$. In Fig. 3 we show a model that is in good 
agreement with CS22949-037. This star has [Fe/H] $\simeq -4.0$ and is also C and N rich$^{10}$, though 
C/Fe and N/Fe ratios are smaller than HE0107-5240. This model is the explosion of a 
30$M_\odot$ star with E$_{51}$ = 20. In this model, the mixing region 
($M_{\rm r} = 2.33 - 8.56 M_\odot$) is 
chosen to be smaller than the entire He core ($M_{\rm r} = 13.1 M_\odot$), in order to reproduce 
relatively large Mg/Fe and Si/Fe ratios. Similar degree of mixing would also 
reproduce the abundances of CS29498-043 (ref. 9), which shows a similar abundance pattern. 
We assume a larger fraction of ejection, 2\%, from the mixed region for CS22949-037 
than HE0107-5240, because the C/Fe and N/Fe ratios are smaller. The ejected Fe mass 
is 0.003 $M_\odot$. The larger explosion energy model is favored for explaining the large 
Zn/Fe, Co/Fe and Ti/Fe ratios$^{14}$. Without mixing, elements produced in the deep 
explosive burning regions, such as Zn, Co, and Ti, would be underproduced. Without 
fallback, the abundance ratios of heavy elements to lighter elements, such as Fe/C, Fe/O, 
and Mg/C would be too large.

In this model, Ti, Co, Zn and Sc are still under-produced, but these 
elements may be enhanced efficiently in the aspherical explosions$^{18}$. 
Almost the same effects as the `mixing and fallback mechanism' are 
realized if the explosion is jet-like, although the total energy can be smaller 
if the beaming angle of the jet is small.
 Similarly, the `mixing and fallback' process can reproduce the abundance 
pattern of the typical EMP stars without enhancement of C and N. For example, the 
averaged abundances of [Fe/H] $\simeq -3.7$ stars in ref.8 can be fitted well with the 
model of 25$M_\odot$ and E$_{51}$ = 20 but with a larger fraction ($\sim$ 10\%) 
of the processed materials in the ejecta. 

In our model, [Fe/H] of several kinds of EMP stars can be 
understood in the supernova-induced star formation scheme$^{19, 20}$. 
In this scheme, [Fe/H] of the second-generation stars are determined 
by the ejected Fe mass divided by 
the mass of hydrogen swept up by the supernova ejecta. As the 
swept-up hydrogen mass is roughly proportional to the explosion energy, 
Fe/H $\propto$ (M(Fe)/0.07$M_\odot$) 
/ E$_{51}$, 
where M(Fe) is the ejected Fe mass. The average stars of [Fe/H] $\simeq -3.7$ (ref. 7), 
CS22949-037, 
and HE0107-5240 correspond to (M(Fe)/0.07$M_\odot$) / E$_{51}$ = 0.07, 0.002, and 0.0004, 
respectively. This correspondence suggests that [Fe/H] of the EMP stars do not reflect 
the age of the stars, but rather the properties of the supernovae, such as the degree of mixing 
and fallback or collimation of a jet.

We have shown that the ejecta of core-collapse supernova explosions of 
$20-130M_\odot$
stars can well account for the abundance pattern of EMP stars. 
In contrast, the observed abundance patterns cannot be explained by
the explosions of more massive, $130-300M_\odot$ stars.
These stars undergo pair-instability supernova and are
disrupted completely$^{14,21}$, which cannot be consistent with the
large C/Fe ratio observed in HE0107-5240 and other C-rich EMP stars.
Also, the abundance ratios of iron-peak elements
([Zn/Fe]$< -0.8$ and [Co/Fe]$< -0.2$) in the ejecta of pair-instability supernovae$^{14,21}$
cannot explain the large Zn/Fe and Co/Fe (ref.14) in the typical 
EMP stars$^{8,22,23}$ and CS22949-037. 
Therefore the supernova progenitors 
that are responsible for the formation of EMP stars are most likely
in the range of $M \sim 20 - 130 M_\odot$, 
but not more massive than 130 $M_\odot$. 

 We have also shown that the most iron-poor star, as well as other C-rich EMP stars, is 
likely to be enriched by massive supernovae that are characterized by relatively small 
Fe ejection. Such supernovae are not hypothetical, but have been actually observed, 
forming a distinct class of type II supernovae (`faint supernovae')$^{24}$. The prototype is 
SN1997D, which is very underluminous and shows quite narrow spectral 
lines$^{7}$: these 
features are well modeled as an explosion of a 25$M_\odot$ star ejecting 
only $2 \times 10^{-3} M_\odot$ 
$^{56}$Ni with small explosion energy E$_{51} \sim 0.4$ (ref. 7). 
SN1999br is a very similar faint 
supernova$^{25}$. On the other hand, typical EMP stars without 
enhancement of C and N 
correspond to the abundance pattern of energetic supernovae 
(`hypernovae'$^{24}$). 
For both cases, black holes more massive than$\sim 3 - 10M_\odot$ must be left as  
results of fallback, suggesting copious formation of the first black holes from 
the first stars. These black holes may make up some of the dark mass in the Galactic halo. 

In our model, HE0107-5240 with [Fe/H]= $-5.3$ was formed from C- and O-enhanced  
gases with [C,O/H] $\sim -1$. With such enhanced C and O, the cooling efficiency is large 
enough to form small-mass stars. In other word, our model predicts
that low-mass EMP stars with [Fe/H] $ < -4$ are likely to have
enhanced [C, N, O/Fe] and [Mg/Fe] in some cases.
A consequence of the low-mass EMP stars being carbon-rich is that 
the population III stars that provided their metals are massive
enough to form (the first) black holes.

\bigskip
\bigskip
{\small
\noindent
1.	Abel, T., Bryan, G. L., \& Norman, M. L. The formation of the first star in the 
Universe. Science 295, 93-98 (2002).

\noindent
2.	Omukai, K. \& Nishi, R. Formation of primordial protostars. Astrophys. J 508, 
141-150 (1998).

\noindent
3.	Nakamura, F. \& Umemura, M. On the mass of population III stars. Astrophys. J 515, 
239-248 (1999).

\noindent
4.	Schneider, R., Ferrara, A., Natarajan, P., \& Omukai, K. First Stars, Very massive 
black Holes, and metals. Astrophys. J. 571, 30-39 (2002).

\noindent
5.	Christlieb, N. et al. A stellar relic from the early Milky Way. Nature 419, 904-906 
(2002).

\noindent
6.	Yoshii, Y. Metal enrichment in the atmospheres of extremely metal-deficient dwarf 
stars by accretion of interstellar matter. Astron. Astrophys. 97, 280-290 (1981).

\noindent
7.	Turatto, M. et al. The peculiar type II supernova 1997D. A case for a very
 low $^{56}$Ni mass. Astrophys. J 498, L129-L133 (1998).

\noindent
8.	Norris, J. E., Ryan, S. G. \& Beers, T. C. Extremely metal-poor stars. VIII.
 High-resolution, high signal-to-noise ratio analysis of five stars with [Fe/H] $<  -3.5$. 
Astrophys. J 561, 1034-1059 (2001).

\noindent
9.	Aoki, W., Ryan, S. G., Beers, T. C. \& Ando, H. The chemical composition of 
carbon-rich, very metal poor stars. Astrophys. J 567, 1166-1182 (2002). 

\noindent
10.	Depagne, E. et al. First stars II. Elemental abundances in the extremely metal-poor 
star CS 22949-037. Astron. Astrophys. 390, 187-198 (2002). 

\noindent
11.	Ryan, S. G. Carbon-rich, extremely metal-poor Population II stars. in CNO in the 
Universe  (eds Charbonnel, C., Schaerer, D., \& Meynet, G.) (in the press); also 
preprint astro-ph/0211608 at $<$http://xxx.lanl.gov$>$. 

\noindent
12.	Fujimoto, M. Y., Ikeda, Y. \& Iben, I. Jr The origin of extremely metal-poor stars 
and the search for population III. Astrophys. J 529, L25-L28 (2000).

\noindent
13.	Siess, L., Livio, M., \& Lattanzio, J. Structure, evolution, and nucleosynthesis of 
primordial stars. Astrophys. J 570, 329-343 (2002).

\noindent
14.	Umeda, H. \& Nomoto, K. Nucleosynthesis of Zinc and iron peak elements in 
population III type II supernovae. Astrophys. J 565, 385-404 (2002).

\noindent
15.	Hachisu, I., Matsuda, T., Nomoto, K., \& Shigeyama, T. Non linear growth of 
Rayleigh-Taylor instabilities and mixing . Astrophys. J 358, L57-L61 (1990).

\noindent
16.	Kifonidis, K., Plewa, T., Janka, H.-Th., \& Muller, E. Nucleosynthesis and clump 
formation in a core-collapse supernova.  Astrophys. J 531, L123-L126 (2000). 


\noindent
17.	Boothroyd, A. I. \& Sackmann, I. -J.  The CNO Isotopes: Deep circulation in red 
giants and first and second dredge-up. Astrophys. J 510, 217-231 (1999).

\noindent
18.	Maeda, K. \& Nomoto, K. Bipolar supernova explosions: nucleosynthesis \& implication
 on abundances in extremely metal-poor stars, preprint
astro-ph/0304172 at $<$http://xxx.lanl.gov$>$ (2003). 

\noindent
19.	Audouze, J. \& Silk, J. The First Generation of Stars: First Steps toward Chemical Evolution of Galaxies. Astrophys. J 451, L49-L52 (1995).

\noindent
20.	Shigeyama, T. \& Tsujimoto, T. Fossil Imprints of the First-Generation Supernova Ejecta in Extremely Metal-deficient Stars. 
Astrophys. J 507, L135-L139 (1998).

\noindent
21.	Heger, A. \& Woosley, S. E. The nucleosynthetic signature of population III.
Astrophys. J 567, 532-543 (2002).

\noindent
22.	McWilliam, A., Preston, G. W., Sneden, C., \& Searle, L. Spectroscopic analysis of 
33 of the most metal poor stars II Astron. J. 109, 2757-2799 (1995).

\noindent
23.	Blake, L. A. J., Ryan, S. G., Norris, J. E., \& Beers, T. C. Neutron-capture elements 
in the Sr-rich, Ba-normal metal-poor giant CS22897-008  Nucl. Phys. A 688, 502-504 (2001).

\noindent
24.	Nomoto, K., Maeda, K., Umeda, H., Ohkubo, T., Deng, J., \& Mazzali, P. 
Hypernovae and their nucleosynthesis. in A massive star odyssey, from main 
sequence to supernova  (eds van der Hucht, K. A., Herrero, A., \& Esteban, C.) (in 
the press); also preprint astro-ph/0209064 at $<$http://xxx.lanl.gov$>$. 

\noindent
25.	Zampieri, L. et al. Peculiar, low luminosity type II supernovae: Low energy 
explosions in massive progenitors?  Mon. Not. R. Astron. Soc. 338, 711-716 (2003).

\bigskip
This work has been supported in part by the grant-in-Aid for Scientific Research of the Ministry of Education, Science, Culture, Sports, and Technology in
Japan.

}

\newpage 

\begin{figure}
\plotone{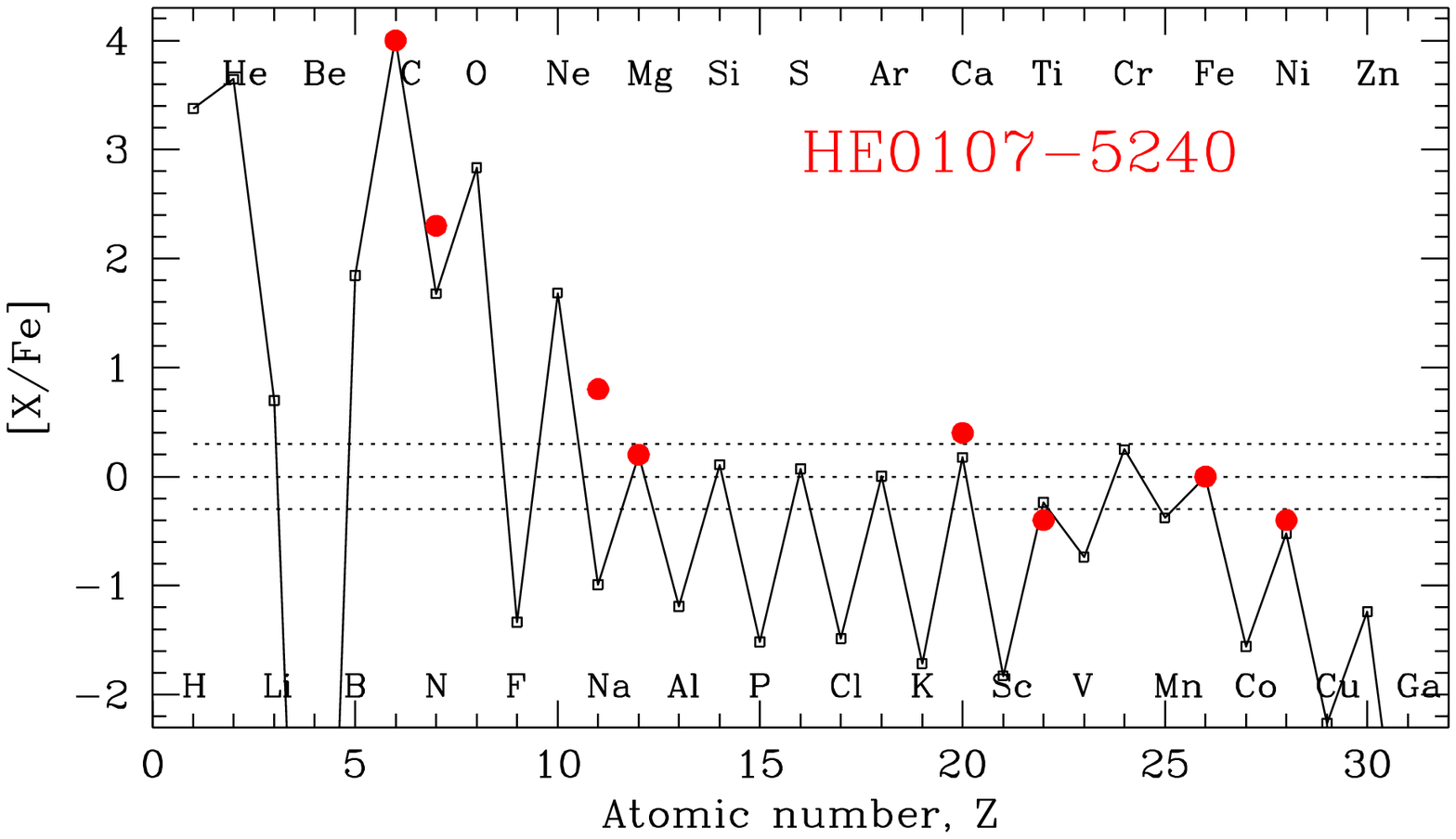}
\caption{Elemental abundances of HE0107-5240, 
compared with a theoretical supernova yield. HE0107-5240 (filled circles)
is the most Fe-deficient, C-rich star yet observed, with [Fe/H]$= -5.3$
and very large ratios of [C/Fe]=4.0 and [N/Fe] =2.3.
Here the supernova 
model is the  population III 25$M_\odot$ core collapse, with relatively 
small explosion energy E$_{51}$ = E$_{\rm exp} 
/10^{51}$ erg =0.3 (see Fig. 2). 
In this model, the supernova ejecta is assumed 
to be mixed inside the He core (within the mass coordinate $M_{\rm r}$= 1.83 and 6.00 $M_\odot$) 
and only a small fraction of the matter, 0.002\% ($f$=0.00002), is ejected 
from this region. The ejected Fe (or $^{56}$Ni) 
mass, $8 \times 10^{-6}M_\odot$, is so small that the large C/Fe 
ratio can be realized. }
\end{figure}

\newpage 

\begin{figure}
\plotone{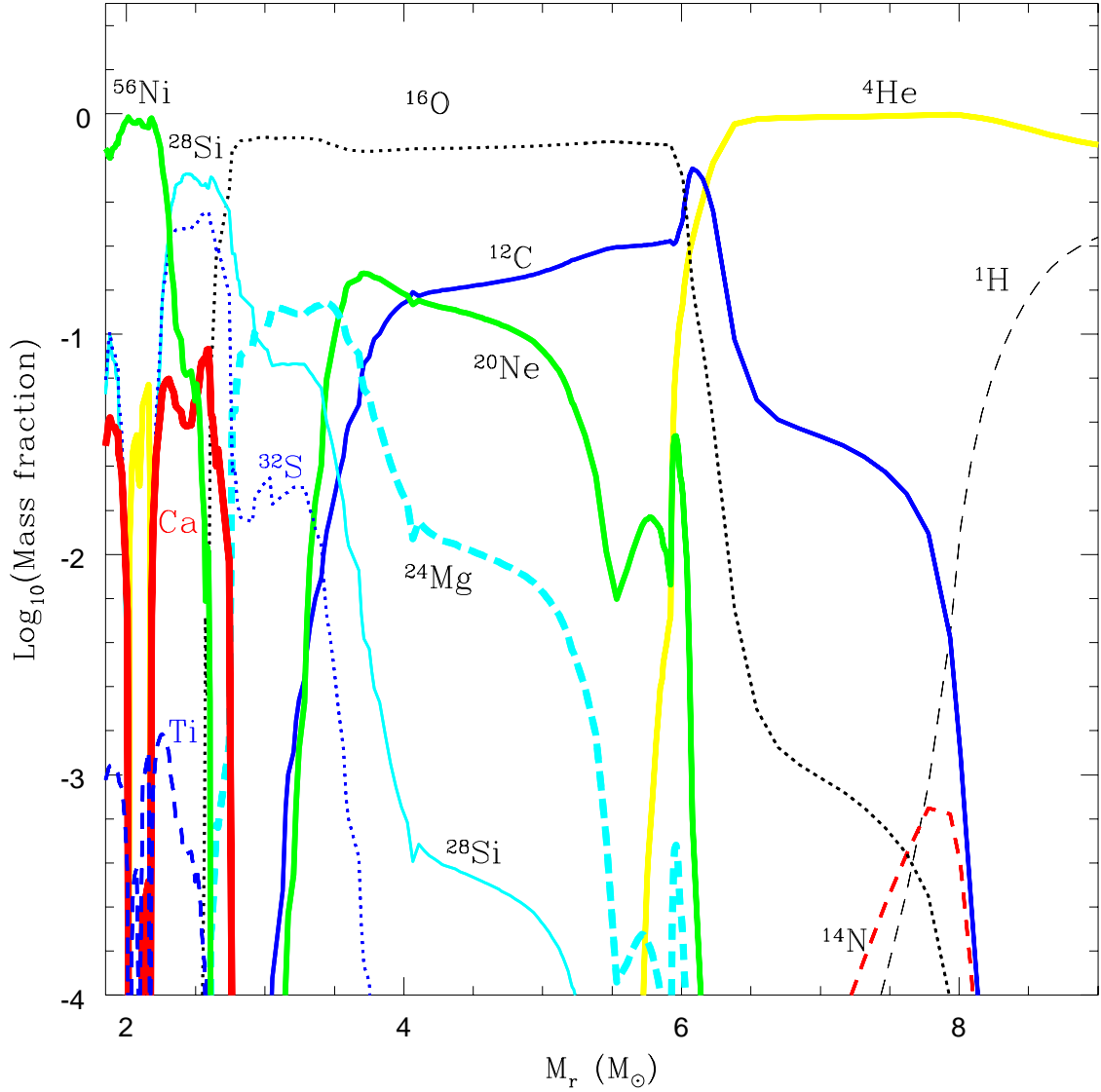}
\caption{The post-explosion abundance distributions for the population III 
25$M_\odot$ model 
with explosion energy E$_{51}$ = 0.3.
Explosive nucleosynthesis takes place behind the shock wave that is 
generated at $M_{\rm r}$ = 1.8 $M_\odot$ and propagates outward. 
The hydrogen-rich envelope
at $M_r >9M_\odot$ is not shown.
}
\end{figure}

\newpage 

\begin{figure}
\plotone{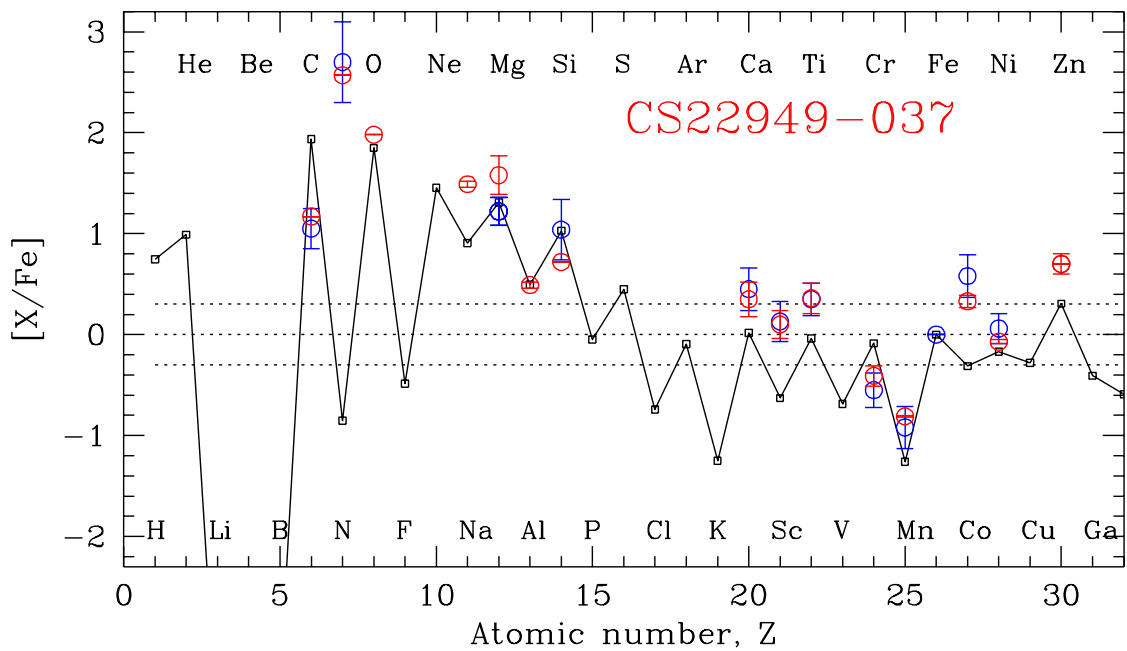}
\caption{Elemental abundances of CS22949-037,
compared with a theoretical supernova yield. The blue and red open circles
are observed data from refs 8 and 10, respectively.
For the theoretical model (open squares and solid lines),
a high energy model (E$_{\rm 51}$ = 20) 
is adopted for the relatively large 
abundances of Co and Zn. 
In this 30$M_\odot$ model, the He core is 13.1 $M_\odot$. The 
mixing region ($M_{\rm r} = 2.33 - 8.56 M_\odot$ )  is smaller than the entire He core, and the 
matter ejection factor f (=0.002) is larger than the model for HE0107-5240 
because of larger Mg/C and C/Fe ratios.} 
\end{figure}

\end{document}